# Controlling Automated Vehicles on Large Lane-free Roundabouts

Mehdi Naderi, Markos Papageorgiou, *Life Fellow, IEEE*, Dimitrios Troullinos,
Iasson Karafyllis, and Ioannis Papamichail

*Abstract*— Controlling automated vehicles on large lane-free roundabouts is challenging because of the geometrical complexity and frequent conflicts among entering, rotating, and exiting vehicles. This paper proposes a comprehensive methodology to control the vehicles within the roundabout and the connected road branches. The developed real-time vehicle movement strategy relies on offline-computed wide overlapping movement corridors, one for each Origin-Destination (OD) movement, which delineate the admissible movement zones of corresponding OD vehicles. Also, space-dependent desired orientations are determined by destination, so as to mitigate potential vehicle conflicts and reduce trip distance. A distributed (per vehicle) movement control strategy, using two nonlinear feedback controllers (NLFC), for circular and straight movements, respectively, is employed to navigate each vehicle within the respective OD corridor toward its destination, accounting for the desired orientation and avoiding collisions with other vehicles; while boundary controllers guarantee that the corridor boundaries will not be violated, and the exit will not be missed. As an overly complicated case study, we consider the famous roundabout of Place Charles de Gaulle in Paris, featuring a width of 38 m and comprising a dozen of bidirectional radial streets, hence a total of 144 ODs. The pertinence and effectiveness of the presented method is verified via microscopic simulation and evaluation of macroscopic data.

## I. INTRODUCTION

To address the issues caused by traffic congestion, such as travel delays, environmental degradation, and decreased traffic safety, various methods of traffic control have been developed and partly employed in the past decades [1], [2]. More recently, a wide variety of Vehicle Automation and Communication Systems (VACS) have been developed that tremendously improve vehicles' individual capabilities, enabling a new generation of potential traffic management tools [3], [4]. This tendency is continuing with the emergence of high-automation or nearly driverless vehicles which are tried out in real traffic environments, see e.g. [5]. In the not-too-far future, vehicles may communicate with each other and with the infrastructure; and drive automatically, based on own sensors, communications, and appropriate movement control strategies.

Recently, the TrafficFluid concept, a novel paradigm for vehicular traffic, which applies at high levels of vehicle automation and communication was proposed [6]. The TrafficFluid concept relies on two combined principles: (a) Lane-free traffic, whereby vehicles are not bound to fixed traffic lanes, as in conventional traffic, but may drive anywhere on the 2-D surface of the road; and (b) Vehicle nudging, whereby vehicles communicate their presence to other vehicles in front of them (or are sensed by them), and this may influence the movement of vehicles in front. Over the last couple of years, several movement strategies were proposed for autonomous vehicles on lane-free infrastructure under the TrafficFluid paradigm, using different methodologies, including: ad-hoc strategies [6], [7], optimal control [8], [9], reinforcement learning [10], nonlinear feedback control [11], [12]; and a generic simulation environment for lane-free traffic has also been developed [13]; see [14] for a brief review.

In a remarkable keynote presentation [15], Luc Julia mentioned two reasons why driverless vehicles may never be a reality, one of them being the intricate Place Charles de Gaulle roundabout in Paris, depicted in Fig. 1, which is too complex for automated vehicles (AV) to navigate. This famous roundabout is 38 m wide, with an outer radius of 84 m and an inner radius of 46 m. It comprises a dozen bi-directional radial streets, i.e., 144 distinct origin-destination (OD) movements for the vehicles. Given this complexity, this road infrastructure is operating without lanes; therefore, once on the roundabout, human drivers must find their way without adhering to any traffic lanes. Luc Julia's statement provided us with the motivation to address the challenge and contemplate the Place Charles de Gaulle roundabout, which is anyhow a lane-free infrastructure, as a case study for the TrafficFluid concept, i.e., to develop a vehicle movement strategy for AV that may populate and drive on such complex roundabouts, as reported in this paper.

Roundabouts are a key element of urban traffic, allowing for more efficient flow at light traffic [17]; but may become a bottleneck point in higher demands. Therefore, successful management of roundabouts, which is considered difficult due to their complexity, may improve the flow of traffic in the surrounding area. There are several works in the literature that focus on AV driving on roundabouts [18]-[33]. A classification of the existing methods, based on some important features, is

\* The research leading to these results has received funding from the European Research Council under the European Union's Horizon 2020 Research and Innovation Programme / ERC Grant Agreement no. 833915, project TrafficFluid, see: https://www.trafficfluid.tuc.gr/.

M. Naderi, M. Papageorgiou, D. Troullinos, and I. Papamichail are with Dynamic Systems and Simulation Laboratory (DSSL), Technical University of Crete, Chania, Crete, Chania, Greece. (e-mail: {mnaderi, mpapageorgiou, dtroullinos, ipapamichail}@tuc.gr); M. Papageorgiou is also with the Faculty of Maritime and Transportation, Ningbo University, Ningbo, China.
I. Karafyllis is with the Dept. of Mathematics, National Technical University of Athens, Athens, Greece. (e-mail: iasonkar@central.ntua.gr)

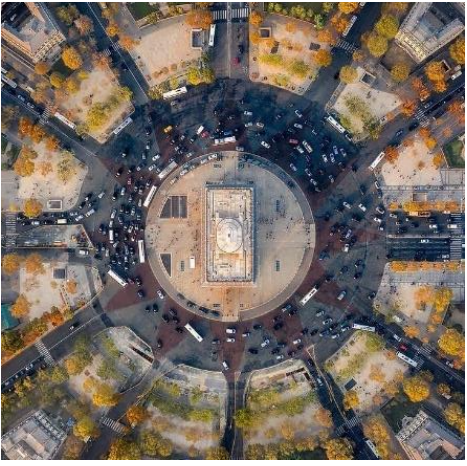

**Fig.1.** Place Charles de Gaulle roundabout [16]

given in Table I. The majority of the reported works appear to focus on simple roundabouts that do not come close to the complexity of this paper's case study. Particularly, most of them concentrate on single or double-lane roundabouts with a limited number of radial streets.

A preliminary report of early results on lane-free roundabouts involving a control scheme for AV and application to the roundabout of Place Charles de Gaulle (Paris) was presented in [34]. Therein, a nonlinear feedback controller, developed in [11] for vehicles moving on straight lane-free roads, was employed to control vehicles on the roundabout. In addition, in [35], we developed an optimal control approach minimizing a weighted sum of the trip distance and the deviation from the circular motion to determine desired orientations on large roundabouts, to replaces the heuristic method used in [34].

In this paper, we extend and improve the strategies presented in [34] in many significant aspects to provide safe and convenient vehicle movements as well as an acceptable throughput, especially in high-density situations. Firstly, a new nonlinear controller, designed for ring-roads in [12], is employed to control vehicles while moving on the roundabout, which is more appropriate than modifying the straight-road controller, as done in [34]. Secondly, a suboptimal online approach, presented in [35], is utilized to determine the desired vehicle orientations. Furthermore, some additional considerations, like adaptive desired speed based on the local density as well as a longitudinal safety controller are introduced to ensure: (i) suitable performance in highly crowded situations; and (ii) good exploitation of the infrastructure and high throughput at all density levels. While designing the movement strategy, we have at parts tried to imagine logical human decisions, and followed them, if they proved efficient. A video of microscopic simulation for the Charles de Gaulle roundabout using the presented approach is available at https://bit.ly/36exR42. Finally, macroscopic data are used to evaluate the traffic-level effectiveness of the presented methodology.

The rest of the paper is as follows. Section II explains the vehicle dynamics and the transformations for circular and skewed movements. The nonlinear controllers used for straight and circular paths are presented in Section III. Section IV describes the designed OD corridors and desired orientation approach. Boundary and safety controllers are presented in Section V. Simulation results are presented in Section VI. Concluding remarks are given in Section VII. Some side issue details are provided in four Appendices.

## II. VEHICLE MODELING

### A. Vehicle Dynamics

The kinematic bicycle model, that has been extensively used in the literature [11], [12], [36], is employed in this study to represent vehicle dynamics. The model variables are visualized in Fig. 2, and the state-space model reads [11]:

$$\begin{aligned} \dot{x} &= v\cos(\theta) \\ \dot{y} &= v\sin(\theta) \\ \dot{\theta} &= \sigma^{-1} v \tan(\delta) \\ \dot{v} &= F \end{aligned} \quad (1)$$

where $x$ and $y$ are the longitudinal and lateral position coordinates of the rear axle midpoint of the vehicle, $v$ is the vehicle speed, and $\theta \in [-\pi/2, \pi/2]$ is its orientation. The model has two control inputs: acceleration $F$ and steering angle $\delta$. Finally, $\sigma$ is the length of the vehicle. To simplify the third state equation above, we define [11]

$$u = \sigma^{-1} v \tan(\delta). \quad (2)$$

Then, (1) can be written as follows:

$$\begin{aligned} \dot{x} &= v\cos(\theta) \\ \dot{y} &= v\sin(\theta) \\ \dot{\theta} &= u \\ \dot{v} &= F \end{aligned} \quad (3)$$

TABLE I. CLASSIFICATION OF REFERENCES ADDRESSING AUTOMATED VEHICLE DRIVING AT ROUNDABOUTS

| | Reference | [18] | [19] | [20] | [21] | [22] | [23] | [24] | [25] | [26] | [27] | [28] | [29] | [30] | [31] | [32] | [33] |
|---|---|---|---|---|---|---|---|---|---|---|---|---|---|---|---|---|---|
| Coverage | Merging only | | * | | | | | | | | | | | | | | |
| | Full Navigation | * | | * | * | * | * | * | * | * | * | * | * | * | * | * | * |
| Model | Double Integrator | * | * | * | * | * | 2D | 2D | * | * | | Real van | * | | | | * |
| | Bicycle Model | | | | | | | | | | * | | | * | * | * | |
| Traffic | Automated Vehicles | * | * | * | * | * | * | * | * | * | * | * | * | * | | * | * |
| | Mixed Traffic | | | | | | | | | | | | | | * | | |
| Roundabout Lanes | | 1 | 1 | 1 | 1 | 1 | 1 | 1 | 1 | 1 | 1 | 2 | 2 | 2 | 2 | 4 | 1 |
| Number of Entries/ Exits | | 4 | 4 | 5 | 3 | 2 | 2 | 4 | 4 | 4 | 4 | 4 | 4 | 4 | 4 | 5 | 3 |
| Merging Lanes | | 1 | 1 | 1 | 1 | 1 | 1 | 1 | 1 | 1 | 1 | 2 | 1 | 2 | 1 | 4 | 1 |
| Control Method | Optimal/ MPC | * | | | * | * | * | * | * | * | * | | * | * | * | | * |
| | Game Theory | | | | | | | | | | | | * | * | * | | |
| | Machine Learning | | | | | | | | | | | | | | | * | * |
| | Heuristic | | * | * | | | | | | | | Fuzzy | | | | | |

## B. Transformation for Circular Movement

While rotating at a roundabout, it is advantageous to transform the above vehicle dynamics to polar coordinates to ease analysis and controller design. Assuming the centre of the roundabout as the origin of the Cartesian and polar coordinates, three new state variables in polar coordinates are defined as below, while the fourth one, i.e. speed, remains unchanged:

$$r = \sqrt{x^2 + y^2}$$
$$\varphi = \tan^{-1}(y/x) \quad (4)$$
$$s = \theta - (\varphi + \pi/2)$$

where $r$ and $\varphi$ are the radius and angle of the vehicle position in polar coordinates, and $s$ is defined as the deviation of vehicle orientation from the circular angle (i.e. from $(\varphi + \pi/2)$). If a vehicle has a completely circular motion, $s$ equals zero; while positive or negative values for $s$ mean, respectively, that the vehicle moves towards or away from the centre. After calculating the time-derivatives of the new state variables, the system dynamics for the transformed model are obtained as

$$\dot{r} = -v\sin(s)$$
$$\dot{\varphi} = v\cos(s)/r \quad (5)$$
$$\dot{s} = u - v\cos(s)/r$$
$$\dot{v} = F$$

## C. Transformation for Skewed Path

In some situations, e.g. moving on entrance or exit branches, a vehicle may have to be guided along a skewed path with angle $\theta' \in [0, 2\pi)$. Transforming the variables to corresponding new coordinates enables employing the controllers originally designed for horizontal roads. The transformation is illustrated in Fig. 3, where the blue lines represent road boundaries in the transformed coordinates, and the skewed coordinates are $(x', y')$ which are derived as below:

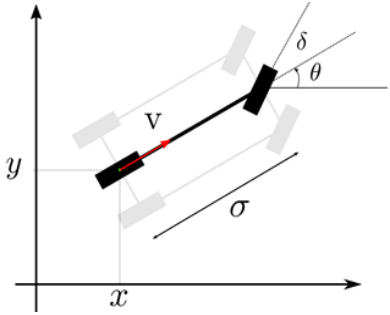

**Fig. 2.** Illustration of the bicycle model variables [11]

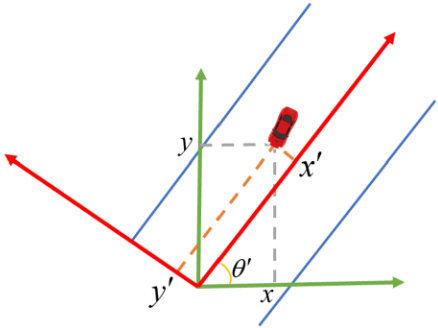

**Fig. 3.** The transformation for skewed path [34]

$$x' = x\cos\theta' + y\sin\theta'$$
$$y' = y\cos\theta' - x\sin\theta' \quad (6)$$

Now, the difference between the vehicle orientation and the skewed angle $\theta'$ is considered as a new state variable $\xi = \theta - \theta'$. Calculating the time-derivatives of the new state variables yields the state equations

$$\dot{x}' = v\cos\xi$$
$$\dot{y}' = v\sin\xi$$
$$\dot{\xi} = u \quad (7)$$
$$\dot{v} = F$$

## D. Sampled-data Bicycle Model

To be implemented in practical frameworks, like simulators, the bicycle model needs to be discretized. Rather than using approximate discretization approaches, like Euler method, the exact sampled-data model is obtained through integrating (1) while considering constant value for control signals during a sampling period. The resulting discrete-time model is [37]:

$$x(k+1) = x(k) + \frac{\sigma}{\tan(\delta(k))} \times$$
$$\left( \sin\left(\theta(k) + v(k)\frac{\tan(\delta(k))}{\sigma}T + F(k)\frac{\tan(\delta(k))}{2\sigma}T^2\right) - \sin(\theta(k)) \right)$$

$$y(k+1) = y(k) + \frac{\sigma}{\tan(\delta(k))} \times$$
$$\left( \cos(\theta(k)) - \cos\left(\theta(k) + v(k)\frac{\tan(\delta(k))}{\sigma}T + F(k)\frac{\tan(\delta(k))}{2\sigma}T^2\right) \right)$$

$$\theta(k+1) = \theta(k) + v(k)\frac{\tan(\delta(k))}{\sigma}T + F(k)\frac{\tan(\delta(k))}{2\sigma}T^2$$

$$v(k+1) = v(k) + F(k)T \quad (8)$$

where $k = 0, 1, 2, \ldots$ is the discrete-time index, and $T$ is the sampling period. For the polar model (5), exact integration appears difficult; hence, we use (8) in the simulation of vehicle dynamics on both straight and circular roads; while the controllers are designed for them based on the respective continuous-time models.

Collision avoidance is the most important goal of control strategies; hence, it is necessary, while simulating vehicle movement, to detect possible collisions among vehicles. Appendix A presents a collision detection procedure for rectangular vehicles.

## III. THE NONLINEAR FEEDBACK CONTROL

Two nonlinear controllers are employed as the kernel for real-time decision making by each vehicle while moving on the roundabout or the connected straight branches, respectively.

### A. Nonlinear Controller for Straight (Horizontal) Roads

A nonlinear feedback controller (NLFC) is developed in [11] to control vehicles on a straight (horizontal) lane-free road that uses the vehicle's state variables and its distance from other adjacent vehicles. This controller was designed for the continuous-time model (1) and guarantees some properties, including avoiding collisions, boundary violation, negative speed, and exceeding the allowable maximum speed. Also, when there is sufficient space, vehicles reach the desired longitudinal speeds, while accelerations, orientations, lateral speeds, and steering angles tend to zero (on an open straight road). The feedback law reads [11]:

$$\delta_i = \tan^{-1}(\sigma u_i / v_i)$$

$$u_i = -\left(v^* + \frac{A}{v_i(\cos(\theta_i) - \cos(\Theta))^2}\right)^{-1} \begin{pmatrix} \mu_1 v_i \sin(\theta_i) + U'(y_i) \\ + p\sum_{j \neq i} V'(d_{i,j}) \frac{(y_i - y_j)}{d_{i,j}} \\ + \sin(\theta_i) F_i \end{pmatrix}$$

$$F_i = -\frac{K_i}{\cos(\theta_i)}(v_i \cos(\theta_i) - v^*) - \frac{1}{\cos(\theta_i)}\sum_{j \neq i} V'(d_{i,j})\frac{(x_i - x_j)}{d_{i,j}}$$

$$K_i = \mu_2 + \frac{1}{v^*}\sum_{j \neq i} V'(d_{i,j})(x_i - x_j)/d_{i,j}$$

$$+ \frac{v_{max}\cos(\theta_i)}{v^*(v_{max}\cos(\theta_i) - v^*)} f\left(-\sum_{j \neq i} V'(d_{i,j})(x_i - x_j)/d_{i,j}\right) \quad (9)$$

where $v^*$ and $v_{max}$ are, respectively, the desired and maximum allowable speeds, and $\mu_1$, $\mu_2$, and $A$ are the controller gains. In particular, $\mu_1$ and $\mu_2$ tune the tracking rate of speed and orientation. The elliptical inter-vehicle distance $d_{i,j}$ is defined as

$$d_{i,j} := \sqrt{(x_i - x_j)^2 + p(y_i - y_j)^2} \quad (10)$$

where $p \geq 1$ is a factor that determines the shape (length versus width) of the considered distance ellipses and is used to adapt the distance metric to the rectangular shape of vehicles. Moreover, $\theta_i$ should remain in $[-\Theta, \Theta]$, where $\Theta \in (0, \pi/2)$ is the maximum allowable deviation of vehicle orientation from the road angle. Also, $V(d_{i,j})$ and $U(y_i)$ are repulsive potential functions for the distance and lateral position, respectively, which are utilized to ensure safe distances from adjacent vehicles and to avoid exceeding the road boundaries, respectively. In this work, the lateral-position potential function is dropped, and its role of avoiding road boundary violation is undertaken by the boundary controllers presented in Section V. On the other hand, the derivative of the used distance potential function is

$$V'(d) = \gamma_1(1/[1 + \exp(\gamma_3 - d/\gamma_2)] - 1) \quad (11)$$

where $\gamma_1$, $\gamma_2$, and $\gamma_3$ are design parameters. Additionally, $f$ is defined as in [11] as

$$f(x) = \frac{1}{2\varepsilon}\begin{cases} 0 & x \leq -\varepsilon \\ (x+\varepsilon)^2 & -\varepsilon < x < 0 \\ \varepsilon^2 + 2\varepsilon x & x \geq 0 \end{cases} \quad (12)$$

where $\varepsilon > 0$ is a constant.

The NLFC (9) was developed for horizontal straight roads, where the desired orientation is zero. However, for vehicles driving on skewed branches, a non-zero desired orientation $\theta_{d,i}$ must be considered. To this end, the controller is modified accordingly by replacing its arguments as follows:
- Replacing the orientation $\theta_i$ with $\theta_i - \theta_{d,i}$.
- Replacing the position coordinates $x_i$ and $y_i$ with $x'_i$ and $y'_i$, respectively, according to (6), considering $\theta' = \theta_{d,i}$.
- Replacing adjacent vehicles' coordinates $x_j$ and $y_j$ with $x'_j$ and $y'_j$, respectively, calculated by (6), considering $\theta' = \theta_{d,i}$.

B. *Nonlinear Controller for Circular Roads*

For vehicle control on the roundabout, we employ an NLFC presented in [12], which is specialized for circular roads, e.g. ring-roads or roundabouts, whose structure is similar to the NLFC for straight roads. It rigorously guarantees (in continuous time) the avoidance of collisions, boundary violation, and exceeding the maximum allowable angular speed, tracking the desired angular speed, convergence of acceleration and orientation (deviation from circular angle) to zero. The feedback law reads:

$$F_i = -K_i(v_i - r_i\omega^*/\cos(s_i)) - (\Phi_i - G_i)r_i\omega^*/\cos(s_i) \quad (13)$$

$$\delta_i = \tan^{-1}\left(\frac{\sigma\cos(s_i)/r_i - }{\sigma(\mu_1\sin(s_i) + (bF_i\sin(s_i) + \Lambda_i)v_i - M_i)/v_i a}\right)$$

where

$$K_i = \mu_2 + \Phi_i - G_i$$
$$+ f\left(-\frac{v_{max}\cos(s_i)}{v_{max}\cos(s_i) - r_i\omega^*}(\Phi_i - G_i)\right)$$

$$\Lambda_i := (v_i\cos(s_i)/r_i - \omega^*)v_i\cos(s_i)/r_i^2 - U'_i(r_i)$$
$$- \sum_{j \neq i}\left(p_{i,j}(r_i - r_j) + r_j(1 - \cos(\varphi_i - \varphi_j))\right)\frac{V'_{i,j}(d_{i,j})}{d_{i,j}}$$

$$a := \left(b - \frac{1}{r_i^2}\right)v_i^2\cos(s_i) + \omega^*\frac{v_i}{r_i} + \frac{A}{(\cos(s_i) - \cos(\Theta))^2} \quad (14)$$

$$\Phi_i := \frac{r_i}{\omega^*}\sum_{j \neq i}V'_{i,j}(d_{i,j})\frac{r_j\sin(\varphi_i - \varphi_j)}{d_{i,j}}$$

$$G_i := \frac{1}{\omega^*}\sum_{j \neq i}\kappa(d_{i,j})\left(g_1\left(\frac{v_j}{r_j}\cos(s_j)\right) - g_1\left(\frac{v_i}{r_i}\cos(s_i)\right)\right)$$

$$M_i := \sum_{j \neq i}\kappa(d_{i,j})\left(g_2(\sin(s_j)) - g_2(\sin(s_i))\right)$$

where $\omega^* \in [0, v_{max}/R_{out}]$ is the desired angular speed, $\mu_1$, $\mu_2$, $A$, and $b > 1/R_{in}^2$ are controller parameters, $\Theta \in (0, \pi/2)$ is the maximum allowable deviation from the circular angle that should satisfy $\cos(\Theta) > R_{out}\omega^*/v_{max}$, and $d_{i,j}$ is the elliptic distance in polar coordinates which is defined as

$$d_{i,j} := \sqrt{p(r_i - r_j)^2 + 2r_i r_j(1 - \cos(\varphi_i - \varphi_j))} \quad (15)$$

where $p \geq 1$ shapes the distance ellipses, which are now curved due to circular movement. As for straight roads, the boundary repulsion function is not needed, since the boundary controllers (Section V) are employed to keep vehicles within the roundabout. Also, functions $f$ and $V(d_{i,j})$ are the ones introduced in (11) and (12). In order to track a desired speed $v^*$ rather than a desired angular speed, $r\omega^*$ is replaced by $v^*$ in the angular speed tracking term of the feedback law, i.e. in $(v - r\omega^*/\cos(s))$.

In the feedback control law, there are two viscous terms, $G_i$ and $M_i$, that aim to reduce the difference between speed and orientation of two vehicles if they are sufficiently close. Since having similar speed is not a goal in this work, we only employ the orientation viscous term, i.e., $M_i$, that helps suppressing collisions in high-density situations. Since the major conflicts mostly happen when the orientations of two vehicles are very different, something that may cause T-bone collisions, reducing the orientation difference, through the orientation viscous term, may mitigate the risk of collision between two vehicles when they are close. The used function for $\kappa(d)$ is:

$$\kappa(d) = \begin{cases} q(\lambda - d)^2 & d < \lambda \\ 0 & d \geq \lambda \end{cases} \quad (16)$$

where $q$ is a positive constant and $\lambda$ is a distance threshold, below which the viscous term is active.

In the above nonlinear controller, as mentioned before, deviation from the circular angle $s$ converges to zero, which means that vehicles tend to have circular motion. However, non-circular movements are also required, e.g. for entering and exiting vehicles; hence, $s$ is replaced with $s - s_{d,i}$ in the feedback law, where $s_{d,i}$ is the desired deviation from the circular angle.

Furthermore, the distance definition, and consequently the iso-distance "aura" surrounding each vehicle, should be changed according to the desired deviation. This calls for a transformation that is described in Appendix B and leads to modified iso-distance curves (aura) around the vehicle, as depicted in Fig. 4, where the curved ellipses surrounding the ego vehicle are aligned with the vehicle's orientation to avoid wasting lateral space around the vehicle while ensuring safety.

It must be emphasized that the iso-distance ellipses extend all around each vehicle. This implies that, based on the NLFCs, "forces" are exerted to each vehicle from obstacles around them. While repulsion from front vehicles is typical in lane-based driving, the NLFCs for lane-free driving also produce "nudging" forces due to obstacles that may be positioned on the left or right or rear side of the vehicle. Beyond safety implications, vehicle nudging was found to have a beneficial impact on the macroscopic properties of the emerging traffic flow [38],[39].

The centre of all mentioned auras, from the ellipses defined for straight or skewed movements to the curved ellipses for the circular movement, is on the middle of the rear axle as introduced in (1). Thus, they cover a smaller area in front of the ego vehicle, compared to the rear part. Having different orientations and moving forward, vehicles have higher collision risk at their front part in crowded situations. To reduce this risk, we move the centre of aura to the middle point of the front axle. Thus, the inter-vehicle distance is calculated based on the position of the middle of the front axle.

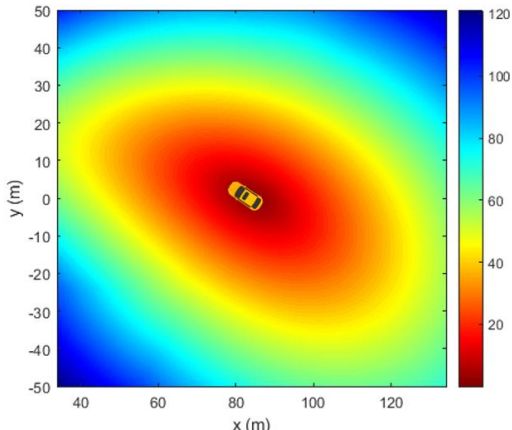

**Fig. 4.** Iso-distance curves in the transformed coordinates

To ensure efficient performance of vehicles, some controller parameters may take different values in different driving phases, i.e. entering, rotating, and exiting. For instance, in the entering and exiting phase, vehicles need strong steering to avoid perpendicular motions or missing the exit, similarly to human driving. Therefore, the corresponding gain $\mu_2$ should be larger than during the rotating phase. Moreover, the parameter $\gamma_1$ of the distance potential function is rendered speed-dependent (like a time-gap) to avoid conservative behavior at low densities.

Finally, the generated control inputs are limited to stay in the ranges $[F_{\min}, F_{\max}]$ and $[-\delta_{\max}, \delta_{\max}]$, respectively.

*C. Adaptive Desired Speed*

The employed NLFCs include a term to track the desired speed or angular speed, which may generate unnecessary accelerations in crowded situations, where high speeds are anyhow not possible. To address this issue, we modify the desired (angular) speed based on the local density $\rho \in [0, 1]$ in the ego vehicle's surrounding area, and Appendix C provides the details of local density computation. Specifically, the adaptive desired speed $v_a^*$ for each vehicle is defined as a function of its local density in a way that corresponds to a triangular fundamental diagram

$$v_a^* = \min\left(v^*, \lambda_s (1/\rho - 1/\rho_{\max})\right) \quad (17)$$

where $\lambda_s$ is a constant and $\rho_{\max}$ is the maximum density calculated by:

$$\rho_{\max} = \sigma w / (\sigma + \sigma_{\text{safety}})(w + w_{\text{safety}}) \quad (18)$$

where $\sigma_{\text{safety}}$ and $w_{\text{safety}}$ are longitudinal and lateral safety distances, respectively. The adaptive desired angular speed $\omega_a^*$ is similarly defied as

$$\omega_a^* = \min\left(\omega^*, \lambda_r (1/\rho - 1/\rho_{\max})\right). \quad (19)$$

Finally, entering vehicles should reduce their speed to avoid conflicts with vehicles rotating on the roundabout close to the corresponding entrance. For this goal, the adaptive desired speed introduced in (17) is modified for these vehicles as

$$v_a^* = \min\left(v^*, \lambda_s (1/\rho - 1/\rho_{\max}), v^*(1 - \rho_{\text{sec}}/\rho_{\max})\right) \quad (20)$$

where $\rho_{\text{sec}}$ is the density of a sector of the roundabout in front of the corresponding entrance. This way, priority is given to the rotating vehicles rather than entering ones. If one is interested in prioritizing the entering vehicles, (20) should be dropped.

IV. OD CORRIDORS AND DESIRED ORIENTATIONS

*A. Defining OD Corridors*

An OD corridor is a part of the roundabout surface, where vehicles with corresponding OD are allowed to drive. In view of the large number of OD couples in big roundabouts, it is sensible to have respective corridors be established automatically, conforming to established rules that come close to human driver decisions. Such corridors can help to mitigate conflicts among vehicles on the roundabout and improve traffic flow. For instance, if a vehicle's destination is close to its entrance branch, it seems logical to avoid driving close to the inner roundabout boundary, which would expose the vehicle to risky and obstructing quasi-perpendicular movement. The roundabout's outer boundary is considered as the outer boundary of all OD corridors, as seen in Fig. 5. Conversely, to achieve better infrastructure utilization, a more pertinent definition of the corridors' interior boundaries is undertaken. First, we categorize ODs into two types: (1) the destination is visible from the origin (Fig. 5(a)); and (2) the destination is not visible from the origin (Fig. 5(b)).

*Visible destination:* For the first type of OD couples where the origin and destination are relatively close to each other and the destination is visible from the origin, the shortest and simplest way to get there is to take a direct path in the vicinity of or on the outer boundary of the roundabout, avoiding excursions to the inner part of the roundabout. In this sense, a simple choice is to consider a straight line connecting the left-most point of the origin branch with the left-most point of the destination branch as the inner corridor boundary, see Fig. 5(a). If such a corridor is too narrow for a certain OD, the inner boundary may be replaced by an arc. For the case of Place Charles de Gaulle roundabout, the destination is visible from the origin if it is up to 3 branches away from the origin. If the exit branch is just after the entrance branch, the second option (arc instead of a line) is used for the inner boundary.

*Invisible destination.* When the destination is not visible from the origin, vehicles cannot drive on a direct path toward the destination and must partially follow a circular motion. In this circumstance, a vehicle is granted the right to gradually utilize the full width of the roundabout until it is sufficiently close to its destination. Then, to avoid missing the exit or moving perpendicularly towards the exit, a line connecting the inner boundary (at the point that has an angular distance of $\varphi_b$ from the exit point) to the left-most point of the exit branch is chosen as the left corridor boundary, as shown in Fig. 5(b).

### B. Specifying Desired Orientations

A vehicle should have some guideline regarding its direction of movement while driving within its OD corridor, so that it first merges in the roundabout traffic, then advances towards its destination and eventually exits. This guideline is provided in the form of desired orientations for the vehicle that are computed based on the vehicle's current position and its destination and are fed to the NLFC to influence the vehicle movement decisions. Thus, in absence of other vehicles, a vehicle would follow the path imposed by the position-depended desired orientations toward its exit. In the presence of other vehicles, the vehicle may have to deviate from that path, e.g. to avoid collision with other vehicles, but will always have a desired orientation corresponding to its current position.

In this work, we employ a weighted average of two orientations, which are the respective optimal solutions of the shortest path to the destination problem; and the minimum deviation from the circular motion problem, see details in [35].

*The shortest path problem:* The shortest path connecting any roundabout position with a specific destination has a clear physical meaning; a vehicle would, in absence of other vehicles, have an interest to drive on the shortest path to its destination. Note, however, that such a path may include strong deviations from the circular angle, which, in presence of other vehicles, increase conflicts with rotating vehicles, causing increased delays and collision risk. Shortest-path orientations are readily derived by distinguishing among two cases:

*Visible* destination: If the straight line connecting the current a position $(r,\varphi)$ with the destination lies completely within the roundabout, then we call the destination "visible"; and the shortest path obviously coincides with that straight line; hence the slope of that line is the desired orientation at $(r,\varphi)$. The visible area for an exit branch, grey-shaded in Fig. 6, is described by

$$\Upsilon = \{(r,\varphi); R_{\text{in}} \leq r \leq R_{\text{out}}, 0 \leq \Delta\varphi \leq \Delta\varphi_{\text{vis}}(r)\} \quad (21)$$

where $\Delta\varphi \in [0, 2\pi)$ is the vehicle's angular distance from the exit point, and $\Delta\varphi_{\text{vis}}(r)$ is a radius-dependent visibility threshold. The visible area is delineated upstream by the inner-circle tangent connected to the exit point, which is displayed light blue in Fig. 6. Using trigonometric relationships, $\Delta\varphi_{\text{vis}}(r)$ is derived as

$$\Delta\varphi_{\text{vis}}(r) = \cos^{-1}(R_{\text{in}}/r) + \Delta\varphi_{\text{vis}}(R_{\text{in}}) \quad (22)$$

where

$$\Delta\varphi_{\text{vis}}(R_{\text{in}}) = \cos^{-1}(R_{\text{in}}/R_{\text{out}}). \quad (23)$$

*Invisible destination*: The shortest path from a roundabout position $(r,\varphi)$ to an invisible destination consists of three parts (Fig. 7). The first part is on the tangent of the inner circle that is connected to the position $(r,\varphi)$, with touch point $(R_{\text{in}}, \varphi + \Delta\varphi_{\text{tan}}(r))$, where $\Delta\varphi_{\text{tan}}(r)$ satisfies

$$\Delta\varphi_{\text{tan}}(r) = \cos^{-1}(R_{\text{in}}/r). \quad (24)$$

The desired orientation in this part is the slope of the tangent. In the second part, the path follows the inner boundary, i.e., the desired orientation is the circular angle, until the destination gets visible; after which we have again the case of visible destination, and the desired orientation is the slope of a line connected to the exit point, see Fig. 7.

In conclusion, the desired orientation initiating a shortest path at every position on the roundabout $(r,\varphi)$, with either visible or invisible destination $(R_{\text{out}}, \varphi_e)$, is:

$$\theta_{\text{d,SP}}(r,\varphi) = \begin{cases} \tan^{-1}(\dfrac{R_{\text{out}}\sin\varphi_e - r\sin\varphi}{R_{\text{out}}\cos\varphi_e - r\cos\varphi}) & 0 \leq \Delta\varphi \leq \Delta\varphi_{\text{vis}}(r) \\ \varphi + \pi/2 & \begin{array}{l}\Delta\varphi > \Delta\varphi_{\text{vis}}(r) \\ \& \; r = R_{\text{in}}\end{array} \\ \tan^{-1}(\dfrac{R_{\text{in}}\sin(\varphi + \Delta\varphi_{\text{tan}}(r)) - r\sin\varphi}{R_{\text{in}}\cos(\varphi + \Delta\varphi_{\text{tan}}(r)) - r\cos\varphi}) & \text{otherwise} \end{cases} \quad (25)$$

and, consequently, the corresponding desired deviation is $s_{\text{d,SP}}(r,\varphi) = \theta_{\text{d,SP}}(r,\varphi) - (\varphi + \pi/2)$. The first condition in (25) refers to positions in the visible area of an exit branch; while the second and third conditions apply when the destination is invisible. Note that the respective tangent slopes, leading to the desired orientations for the first and third conditions, are calculated after transforming the respective two points' positions to Cartesian coordinates, as the ratio $\Delta y/\Delta x$ of their difference in $y$ coordinate ($\Delta y$) over their difference in $x$ coordinate ($\Delta x$).

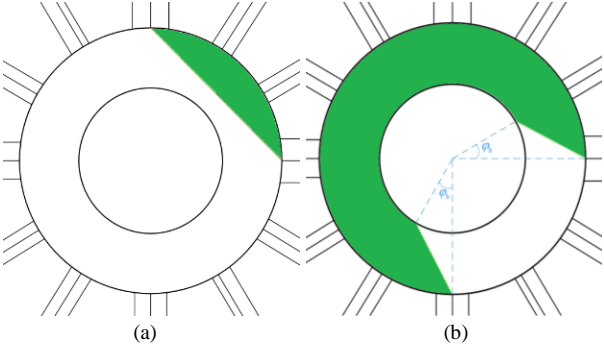

**Fig. 5.** Defined corridors: (a) visible ODs, (b) invisible ODs

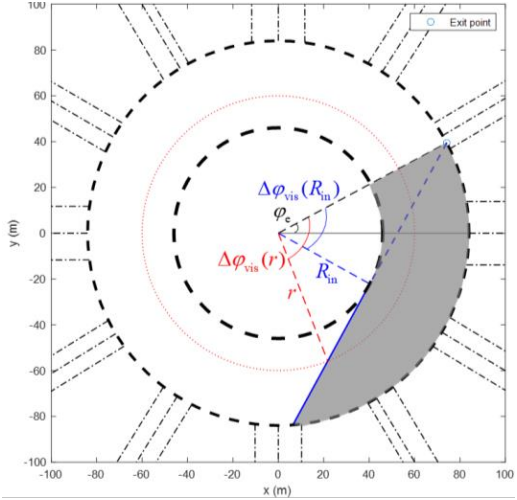

**Fig. 6.** The visible area (grey-shaded) for an exit point

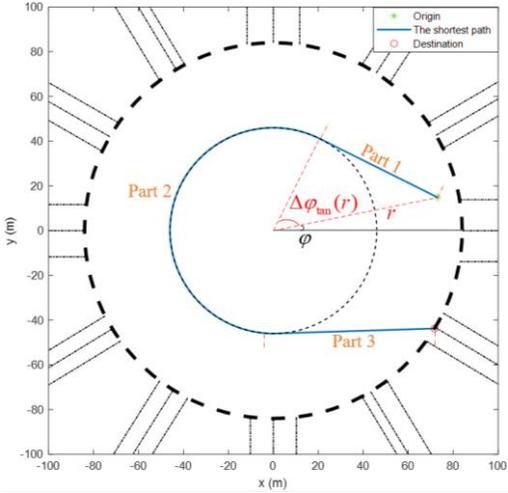

**Fig. 7.** The shortest path for an invisible destination

*The minimum deviation problem:* A path connecting any position in the roundabout with a destination with minimum deviation from circular motion is interesting because most vehicles are rotating and if their orientations are close to the circular angle, then they are close to each other, something that mitigates vehicle conflicts and the strength of any required collision-avoidance maneuvers. In [35], the derived solution of an optimal control problem indicates that deviations from the circular angle are minimized if a vehicle retains a constant deviation on its path from any position to the destination and this constant deviation is

$$s_{d,MD}(r,\varphi) = \tan^{-1}\left(-\frac{\ln(R_{out}) - \ln(r)}{\varphi_e - \varphi}\right). \quad (26)$$

Clearly, the minimum-deviation path for a vehicle that has just entered the roundabout, i.e., at $r = R_{out}$, is to move along the outer boundary, as that results in zero deviation, and is obviously undesirable since all vehicles would move on the boundary or close to it while the inner roundabout parts would remain unexploited.

Having the solutions of both extreme choices, namely the shortest path and the minimum-deviation path, each with its strengths and weaknesses, the desired orientation for a vehicle moving on the roundabout is determined at each position $(r,\varphi)$ by a convex combination of both

$$s_d(r,\varphi) = \alpha s_{d,SP}(r,\varphi) + (1-\alpha)s_{d,MD}(r,\varphi) \quad (27)$$

where $0 \leq \alpha \leq 1$ can be utilized to appropriately distribute vehicles on the roundabout surface. In particular, by choosing a small value for this parameter, vehicles tend towards minimum-deviation paths which causes a high exploitation level of the outer roundabout parts. On the other hand, a large value of $\alpha$ navigates many vehicles toward the inner parts. Hence, we may choose a medium value to have a desired distribution of the vehicles on the roundabout. An additional possibility is to choose a random value of $\alpha$, different for each entering vehicle, from an appropriate interval, attempting to reach a more uniform vehicle distribution.

## V. BOUNDARY AND SAFETY CONTROLLERS

### A. Lateral Boundary Controllers

Ensuring that a vehicle will not go beyond the road boundaries is of greater importance in a lane-free traffic environment than it is in one with lanes where vehicles tend to drive in the middle of each lane, significantly reducing the risk of going off the road. Furthermore, in dense traffic conditions of lane-free traffic, it is beneficial to have some vehicles driving exactly on a (left or right) road boundary, of course without ever violating it, because this increases the exploitation level of the available road width. Note that boundary controllers are also employed for the corridors of Section IV, whose boundaries must also be respected, even if they do not coincide with the road boundaries.

To this end, we deploy two boundary controllers, one for each side of the road or corridor, which provide upper and lower limits for the steering angle generated by the NLFC of the vehicle. The vehicle drives according to NLFC if such a bound is not activated; if one of the bounds is steadily activated, the vehicle is asymptotically navigated, by the action of the respective boundary controller, to the corresponding road boundary and continues driving exactly on the road boundary for as long as the steering angle produced by NLFC is activating that bound, e.g. due to sideward nudging by other vehicles. The boundary controller is first designed to handle a straight road boundary; and is eventually adapted to also accommodate circular and skewed paths.

*Boundary controller for straight (horizontal) roads:* If a boundary controller is activated, vehicles will asymptotically reach to a road boundary, whereby overshooting is not allowed, since it would imply boundary violation. To design a boundary controller for a straight road, we consider the bicycle model (1) and have two main goals: convergence of the vehicle's lateral position $y(t)$ to the road boundary; and convergence of the vehicle orientation $\theta(t)$ to zero. Due to the minor effect of vehicle's longitudinal position and speed in these endeavors, a subsystem of the bicycle model, including only the two state variables of interest, $y(t)$ and $\theta(t)$, and one control variable, $\delta(t)$, is considered; while the speed $v$, that appears in the subsystem equations, may be considered as a measurable exogenous quantity (disturbance).

A simple but efficient approach for designing a boundary controller involves linearizing the subsystem around the desired equilibrium point ($y = y_d, \theta = 0, \delta = 0$) and employing a linear state-feedback controller where $y_d$ is the desired lateral position, i.e., the road boundary. Then, the linearized system reads

$$\begin{bmatrix} \dot{y}(t) & \dot{\theta}(t) \end{bmatrix}^T = \mathbf{A}\begin{bmatrix} y(t) - y_d & \theta(t) \end{bmatrix}^T + \mathbf{B}u(t) \qquad (28)$$

with

$$\mathbf{A} = \begin{bmatrix} 0 & v^* \\ 0 & 0 \end{bmatrix}, \mathbf{B} = \begin{bmatrix} 0 \\ 1 \end{bmatrix}.$$

Then, the following linear state-feedback controller can be developed for the subsystem:

$$u(t) = -\mathbf{K}\begin{bmatrix} y(t) - y_d & \theta(t) \end{bmatrix}^T \qquad (29)$$

where $\mathbf{K}$ is the feedback gain, which can be readily computed to have real and negative closed-loop eigenvalues that provide asymptotic convergence of the state variables to their desired values.

*Boundary controller for circular roads:* As per Section IV, at least one of the boundaries considered for a vehicle moving on an OD corridor is circular; therefore, a boundary controller should be developed for circular movements as well. Following the same design procedure presented for the straight road, a subsystem containing vehicle's radius $r$ and deviation from the circular angle $s$ is considered for the design of the boundary controllers. The system is linearized at the equilibrium point ($r = r_d$, $s = 0$, $\delta = \delta_d$), while the external variables for the subsystem are set $v = v^*$ and $F = 0$; where $r_d$ is the desired radius, i.e., the inner or outer radius of the roundabout, and it is easy to show that the stationary value of the steering angle $\delta_d$ in circular motion with radius $r_d$ is given by

$$\delta_d(r_d) = \tan^{-1}(\sigma/r_d). \qquad (30)$$

Given (5), the linearized subsystem reads

$$\begin{bmatrix} \dot{r}(t) & \dot{s}(t) \end{bmatrix}^T = \mathbf{A}\begin{bmatrix} r(t) - r_d & s(t) \end{bmatrix}^T + \mathbf{B}(u(t) - u_d(r_d)) \qquad (31)$$

with

$$\mathbf{A} = \begin{bmatrix} 0 & -v^* \\ v^*/r_d^2 & 0 \end{bmatrix}, \mathbf{B} = \begin{bmatrix} 0 \\ 1 \end{bmatrix}, \ u_d(r_d) = \sigma^{-1}v\tan(\delta_d(r_d)). \qquad (32)$$

Then, considering (2), (31) and (32), the controller is

$$u(t) = -\mathbf{K}_c\begin{bmatrix} r(t) - r_d & \theta(t) - \theta_r(t) \end{bmatrix}^T + v(t)/r_d \qquad (33)$$

where $\mathbf{K}_c$ is the feedback gain for the circular boundary controllers.

*Boundary controller for skewed paths.* For skewed boundaries, the corresponding model was presented in Section II.C, and the skewed boundary controller reads

$$u(t) = -\mathbf{K}\begin{bmatrix} y'(t) - y'_d & \xi(t) \end{bmatrix}^T \qquad (34)$$

where $y'_d$ is the desired lateral position for the vehicle in the transformed coordinates, and $\mathbf{K}$ is the same feedback gain introduced for the straight boundary controller.

The boundary controllers (29), (33), (34) are applied to corresponding types of corridor boundaries to produce for each vehicle lower (left boundary) and upper (right boundary) bounds for the steering angle, denoted $\underline{\delta}$ and $\bar{\delta}$, respectively.

### B. Longitudinal Safety Controller

Since several modifications were introduced to the original NLFCs, such as different desired orientations, practical limits for the acceleration and steering angle and discrete-time implementation, the original controllers' features, e.g. avoiding collisions, cannot be strictly guaranteed anymore. Consequently, vehicle collisions might occur in some rare severe circumstances. To prevent such situations, we develop a safety controller, imitating human drivers' behavior, generating a safe upper limit for vehicle acceleration.

The general procedure pursued is summarized as follows:
(i) Consider the obstacles in close vicinity of the ego vehicle (EV).
(ii) Estimate a possible conflict position for each obstacle.
(iii) Select the conflict position with the shortest Euclidean distance to the EV.
(iv) Let a linear state-feedback controller determine a safe upper limit for the EV acceleration to ensure that no collision will occur.

Regarding (i), only obstacles located in a limited vicinity of the EV, i.e. satisfying $D_{i,j} < D_{th}(v)$, are considered; where $D_{i,j}$ is the Euclidean distance between two vehicles $i$ and $j$ (i.e., the EV and the obstacle), and $D_{th}(v) = D_0 + D_1 v$, with positive constants $D_0$ and $D_1$, is a speed-dependent threshold (time-gap) to avoid conservative behavior at low speeds.

Regarding (ii), a rough prediction of the short-term future path of the EV and each obstacle is performed in order to predict a possible conflict, i.e., crossing of their paths. To this end, we classify each vehicle motion into circular or skewed types, based on the value of its desired orientation. Specifically, if the desired deviation from circular angle is less than a threshold, e.g. $10°$, the motion is classified as circular, that means the vehicle will proceed on its current radius. Otherwise, the motion is classified as skewed, i.e., the vehicle will move along its desired orientation. After this classification, cross-points of future paths of the EV and each obstacle are determined based on the four possible notion combinations:

*Both EV and obstacle have circular motions*: There is no potential cross-point in this condition since both vehicles move in parallel.

*Both EV and obstacle have skewed motions*: In this case, a single cross-point is calculated by crossing the two vehicle motion equations:

$$y = \tan\theta_{d,i}(x - x_i) + y_i$$
$$y = \tan\theta_{d,j}(x - x_j) + y_j \quad (35)$$

Then, if their desired orientations are not identical (in which case their paths do not cross), the cross-point is

$$x_c = \frac{y_j - y_i + x_i \tan\theta_{d,i} - x_j \tan\theta_{d,j}}{\tan\theta_{d,i} - \tan\theta_{d,j}}$$
$$y_c = \tan\theta_{d,i}(x_c - x_i) + y_i \quad (36)$$

If the cross-point is out of the roundabout or behind the EV or the obstacle in the skewed coordinates aligned with the EV's desired orientation, it is ignored. Otherwise, it is considered as a candidate for the closest cross-point.

*EV and obstacle have circular and skewed motions, respectively*: In this condition, computation of a possible cross-point is slightly more elaborated and provided in Appendix D.

*EV and obstacle have skewed and circular motions, respectively*: See Appendix D for cross-point specification.

In addition, the closest obstacle located in a narrow corridor around the estimated future path of the ego vehicle is identified (see Appendix D for details), and its current position is considered as a possible conflict location.

Regarding (iii), among all estimated cross-points and the closest obstacle position in front of the EV, the closest one is taken as the conflict point, to which the EV should maintain a safety distance by the action of the safety controller.

Finally, regarding (iv), a safety controller is designed, considering simple double-integrator dynamics to reflect the approach of the EV to the conflict point

$$\begin{bmatrix} \dot{D}_o \\ \dot{v} \end{bmatrix} = \begin{bmatrix} 0 & -1 \\ 0 & 0 \end{bmatrix} \begin{bmatrix} D_o \\ v \end{bmatrix} + \begin{bmatrix} 0 \\ 1 \end{bmatrix} F_s \quad (37)$$

where $D_o$ is the distance between the EV and the conflict point and $F_s$ is the safety acceleration to be imposed as an upper limit for EV's acceleration. The state feedback reads

$$F_s = \mathbf{K}_s [D_o - D_s \quad v]^T \quad (38)$$

where $D_s$ is the safety distance and $\mathbf{K}_s$ is the feedback gain chosen such that the closed-loop eigenvalues are real and negative to ensure that the distance will not get smaller than the safety distance.

To sum up, the structure of the overall proposed control strategy, including the nonlinear, boundary and safety controllers, is illustrated as Fig. 8.

## VI. SIMULATION RESULTS

The developed control strategy is generic and can be applied to any complex circular roundabout. In fact, in the design procedure, we do not use any specific details of the case study. The only geometrical assumption is that inner and outer boundaries are circular, which is shared by a large number of roundabouts. Also, the branches are introduced by angular positions and widths; hence, any possible topology of branches, either symmetric or asymmetric, can be considered. In this section, we consider, as a case study, the famous Place Charles de Gaulle roundabout (Paris, France), that has an outer (inner) radius of 84 m (46 m), hence featuring a width of 38 m. It comprises 12 bidirectional branches, resulting in 144 distinct OD movements. The horizontal branch on the right in Fig. 1 is the famous Champs-Élysées Avenue. This is likely the most complex real case study for a roundabout.

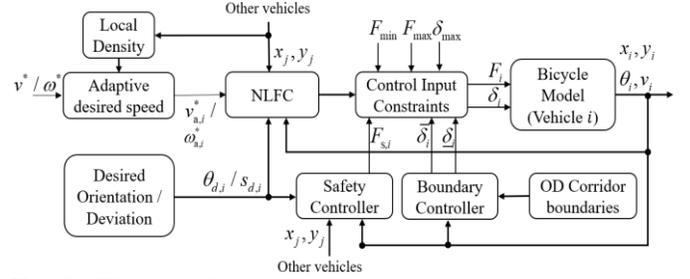

**Fig. 8.** The control system structure

### A. Simulation Set-up and Control Parameters

The proposed control system is implemented in the TrafficFluid-Sim platform [40] which is an extension of the SUMO simulator [41]. Several modifications have been made to facilitate the simulation of vehicles on a roundabout using the proposed strategy [40]. In particular, the discrete-time bicycle model (8) was added to the simulator to represent vehicle dynamics with higher accuracy than the double-integrator model, in view of stronger vehicle turnings in the urban environment. All 144 OD routes are considered, and the individual OD demand flows are defined proportional to the product of the widths of the corresponding entrance and exit branches. The simulation covers, beyond the roundabout, also the connected branches up to a distance of 65 m from the outer roundabout boundary. Thus, entering (exiting) vehicles are released (dropped) at 65 m upstream (downstream) of the roundabout.

The sample period for vehicle dynamics and controllers is 0.1 s, emulating real-time operation of the suggested approach. Furthermore, we have: length and width of vehicles $\sigma = 4.2$ m and $w = 1.7$ m, respectively; lower and upper limits of acceleration $F_{\min} = -4$ m/s$^2$ and $F_{\max} = 0.6$ m/s$^2$, respectively; maximum steering angle $\delta_{\max} = 50°$; desired speeds $v^* = 12$ m/s and $\omega^* = 0.143$ rad/s. Further parameter values are given in Appendix E.

### B. Microscopic Results

To assess the pertinence of the proposed methodology, a simulation of 12 vehicles with various ODs on Place Charles de Gaulle roundabout is initially conducted. The simulation results are depicted in Figs. 9 - 12. As can be seen in Fig. 9, vehicle paths on the roundabout are smooth, and all vehicles are appropriately guided to their respective destinations. The vehicles with distant destinations, e.g. vehicles 1, 4, and 11, drive partially close to the inner boundary. Other vehicles, particularly those with closer destinations, like vehicles 5, 6, and 12, do not approach the inner roundabout parts.

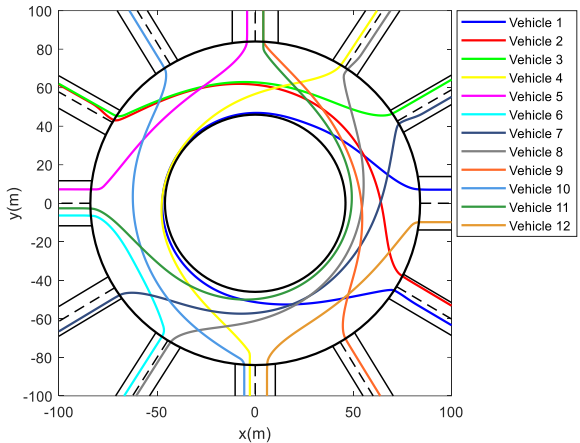

**Fig. 9.** Vehicle paths for 20 vehicles

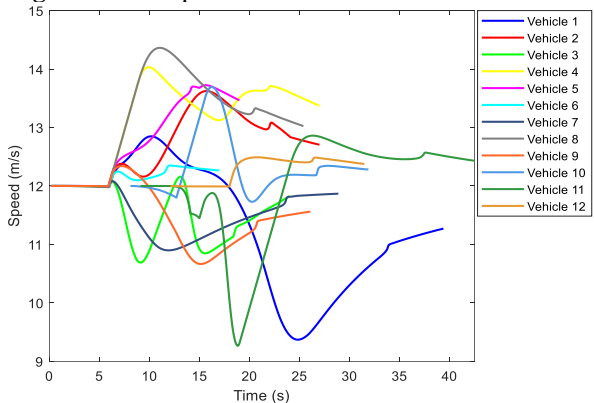

**Fig. 10.** Vehicle speeds for 20 vehicles

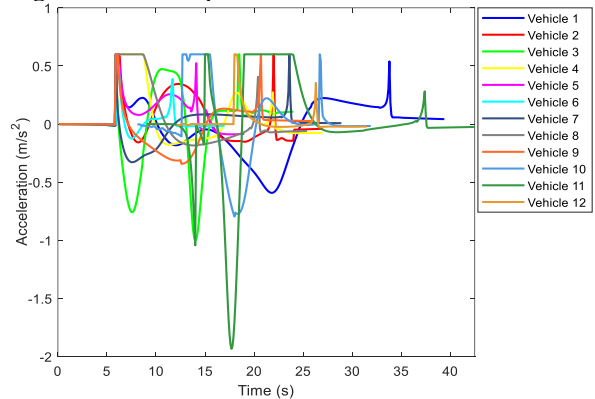

**Fig. 11.** Vehicle accelerations for 20 vehicles

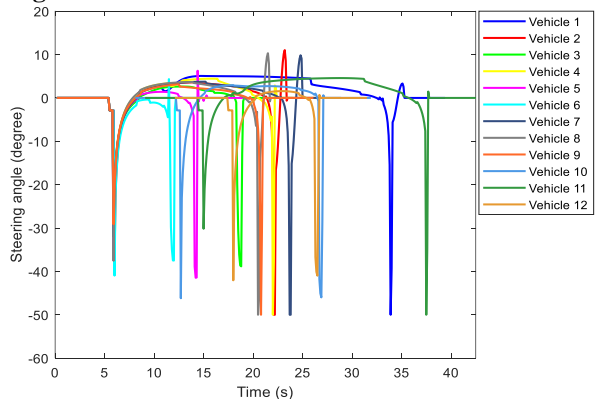

**Fig. 12.** Vehicle steering angles for 20 vehicles

Vehicles tend to follow the desired speed, i.e. 12 m/s, both on the branches and on the roundabout. However, they may temporarily deviate (positively or negatively) from the desired speed to avoid collisions, see Fig. 10. For instance, vehicle 8 is nudged by other vehicles and reaches up to 14.5 m/s, which is still less than the maximum allowed speed. In addition, Fig. 11 illustrates vehicle accelerations in which their efforts to, on one hand, track the desired speed and, on the other hand, avoid colliding with other vehicles via occasional decelerations can be observed. Note that, when the roundabout is not crowded, like in this simulation, vehicles drive fast, and some of them, e.g. vehicles 3 and 11, may need to generate strong deceleration when encountering another vehicle approaching from a different angle. In contrast, in high-density situations, vehicles drive at lower speed, and consequently any produced collision-avoidance decelerations are smaller.

Finally, vehicle steering angles are displayed in Fig. 12 where sharp values reasonably occur when a vehicle enters or exits from the roundabout. In fact, it is necessary to have a significant turning, similar to human driving, when entering the roundabout, to avoid perpendicular motion. Also, while exiting, vehicles need to turn relatively sharply to adapt to the exiting branch orientation. Besides, some smooth steering angles are generated while rotating on the roundabout to avoid collisions.

A second simulation involving a total of 1600 vehicles over a period of 26 min is carried out to confirm the effectiveness of the presented approach in crowded situations. A snapshot of the simulation is depicted in Fig. 13, and a video is available at https://bit.ly/36exR42. As shown in the video, vehicles move towards their destinations without exceeding the roundabout boundaries, though occasionally driving on them by the intervention of the boundary controller, and without any collision. When the roundabout is not crowded (at the start and end of the simulation), vehicles drive fast, close to their desired speed; slowing down occasionally to give way to other vehicles, as in the 12-veh simulation above. In crowded situations, vehicles may have frequent limited accelerations and decelerations to advance in the traffic and avoid collisions, as also experienced in a human-driven vehicle in similar conditions.

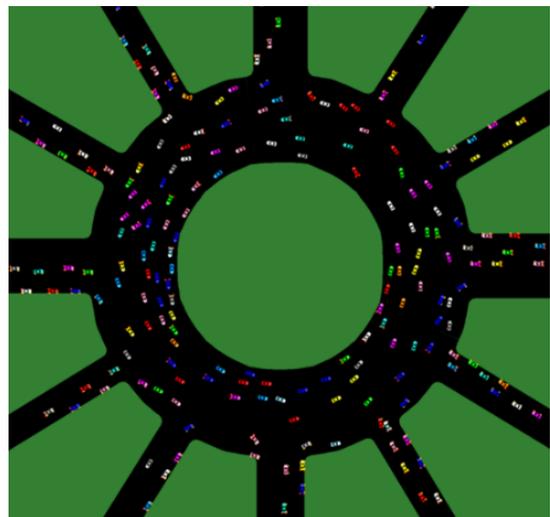

**Fig. 13.** Snapshot of microscopic simulation for the Charles-de-Gaulle roundabout

*C. Macroscopic Results*

As mentioned earlier, roundabouts are critical elements within urban road networks, hence it is important that the employed vehicle movement strategy yields an appropriate throughput. To investigate such traffic flow aspects, the extraction of macroscopic traffic data of the roundabout operation is useful. Specifically, considering a sampling period of 1 min, we measure a few variables, including the number of vehicles on the roundabout, the number of exited vehicles, the average speed of vehicles on the roundabout, and the roundabout average flow; the last one is the average of 12 detector measurements installed between each two consecutive branches.

The roundabout's fundamental diagram (FD) is obtained by plotting the number of exited vehicles or the average flow on the roundabout versus the number of vehicles on the roundabout (density) for all sampling periods, as illustrated in Fig. 14 for the case corresponding to the presented video. It may be seen that, as the demand arriving from the branches (and hence the roundabout density) increases, the throughput on the roundabout increases, until it saturates for high density values. No throughput deterioration is observed at high densities. This macroscopic behavior appears because of giving priority to the rotating vehicles, since the entering vehicles reduce their speed and merge into the roundabout more conservatively if they encounter high density (see Section III.C). As a result, although the average speed reduces at higher densities, no gridlock, the situation in which the average speed drops to almost zero, appears on the roundabout, while some small queues can be observed on the entrance branches.

The time-histograms of the mentioned macroscopic variables are shown in Fig. 15 where the minimum average speed occurs when the number of vehicles on the roundabout is maximum. Note that the decrease in the flow and the number of exited vehicles, after the 19$^{th}$ sample, is due to the smaller entrance flows.

It is interesting to investigate what happens if the entering vehicles, rather than the rotating ones, are prioritized. This can be implemented by making some minor changes in the control system. Firstly, the dependence of the desired speed of the entering vehicles on the density of the roundabout sector, i.e. (20), is canceled. Secondly, we increase the potential function of entering vehicles, as seen by rotating vehicles, to increase their importance (repulsion) in collision avoidance. As a result, in conflicts between an entering vehicle and a rotating one, the latter yields to let the former enter the roundabout. Fig. 16 depicts the resulting FD where now the decreasing flow part becomes visible at high densities. In fact, the vehicles continue entering the roundabout even if there are many vehicles in front of them, which results in quasi-gridlock, while there is no queue on the entrance branches. In this situation, the average speed decreases to values close to zero, while the number of exited vehicles and average flow drop significantly, see Fig. 17. Notice that the change affects the FD mainly for high densities, while the capacity and critical density differences are minor. Interestingly, in the real system, entering vehicles have priority over rotating vehicles on the roundabout. Unfortunately, it was not possible to receive real data reflecting the current operation of the roundabout.

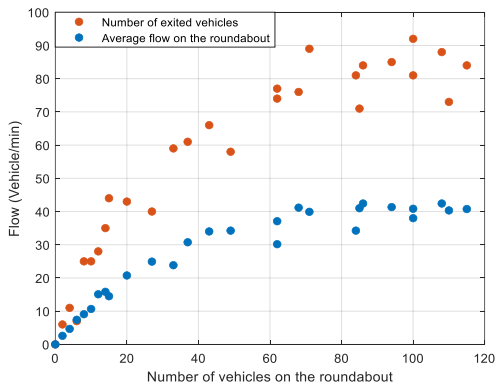

**Fig. 14.** The FD with priority of rotating vehicles

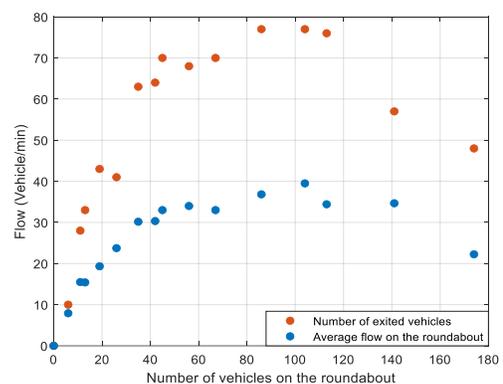

**Fig. 16.** The FD with priority of entering vehicles

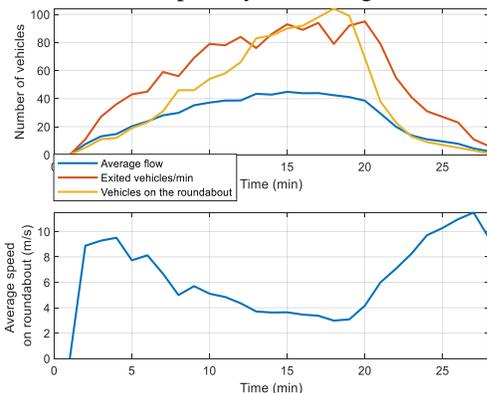

**Fig. 15.** The macroscopic variables with priority of rotating vehicles

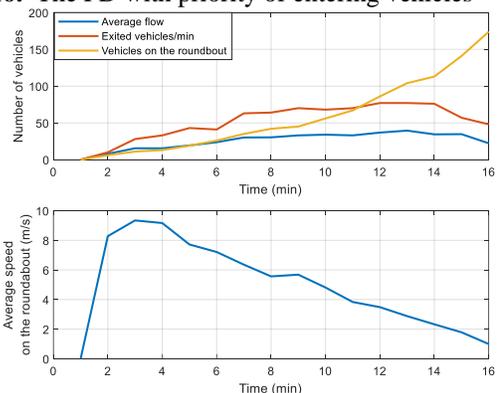

**Fig. 17.** The macroscopic variables with priority of entering vehicles

## VII. CONCLUSION

This paper proposes a comprehensive and efficient control scheme for vehicles moving on large lane-free roundabouts. The vehicle dynamics are represented by the bicycle model and its respective transformations for skewed and circular movements. Two nonlinear controllers are employed as the kernel of the vehicle movement strategies on the roundabout or the connected branches. Some modifications are introduced to facilitate entering and exiting maneuvers and to ensure performance in high-density situations. Also, a weighted average of optimal solutions of the shortest-path and minimum-deviation problems is considered as the desired deviation to be fed to the circular controller. Furthermore, linear state-feedback-based boundary and safety controllers are designed to avoid boundary violation or collisions in severe conditions. The simulation results, including microscopic and macroscopic data, confirm the effectiveness of the presented method in different conditions and show its flexibility in implementing various policies, like prioritizing entering vehicles.

The suggested approach was designed so as to fulfill a number of significant requirements, such as collision avoidance, passenger convenience, boundary respect, guaranteed vehicle's exit at their respective destinations, no roundabout blocking; while yielding a reasonably high throughput. Naturally there may be also other ways to control vehicles on lane-free roundabouts that meet these goals and possibly lead to higher throughput. Ongoing work is investigating such possibilities.

## APPENDIX A: COLLISION DETECTION

Detecting collisions among vehicles is essential, while simulating vehicle movement. A sufficient condition for no-collision is derived by specifying a threshold for the Euclidean distance $D_{i,j}$ between two vehicles $i$ and $j$, above which there is certainly no collision between them. Assuming a rectangle with length $\sigma$ and width $w$ as the shape of each vehicle, such a threshold is determined by considering the situation with maximum Euclidean distance in which two vehicles collide, as shown in Fig. 18, where the distance between two vehicles is $\sqrt{4\sigma^2 + w^2}$. Consequently, a sufficient condition to ensure that no collision occurs is:

$$D_{i,j} > D_{NC} = \sqrt{4\sigma^2 + w^2}; \ \forall i, j; \ i \neq j. \quad (39)$$

Since there are collision-free situations that do not satisfy (39), we employ, if $D_{i,j} \leq D_{NC}$, an exact (but computationally more expensive) collision detection approach which uses vehicle positions and orientations and checks whether at least one vertex of one of the two vehicles is within the other vehicle, as shown in Fig. 19. First, the position of vertices of each vehicle are calculated

$$\begin{aligned} V_1 &= (x + w\sin\theta/2, y - w\cos\theta/2) \\ V_2 &= (x - w\sin\theta/2, y + w\cos\theta/2) \\ V_3 &= (x + w\sin\theta/2 + \sigma\cos\theta, y - w\cos\theta/2 + \sigma\sin\theta) \\ V_4 &= (x - w\sin\theta/2 + \sigma\cos\theta, y + w\cos\theta/2 + \sigma\sin\theta) \end{aligned} \quad (40)$$

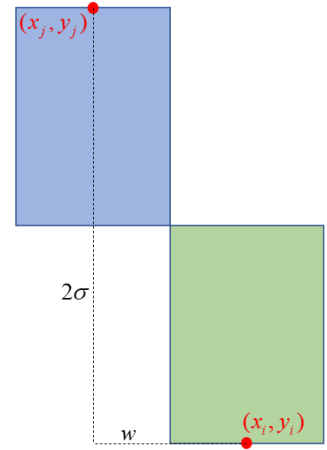

**Fig. 18.** Collisions with the maximum distance

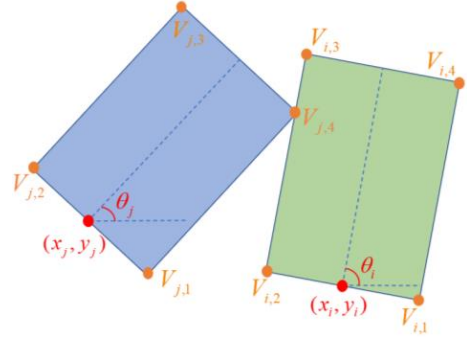

**Fig. 19.** Exact collision detection details

The next step is to check if any vertex of vehicle $j$ is included in the rectangle of vehicle $i$. To this end, the vertices of vehicles $i$ and $j$ are transformed in new coordinates with origin the vehicle $i$ position $(x_i, y_i)$ and $x_i$'-axis along its orientation $\theta_i$. In the new coordinates, the rectangle of vehicle $i$ comprises the points $\{(x, y); 0 \leq x_i' \leq \sigma, -w/2 \leq y_i' \leq w/2\}$, and a vertex of vehicle $j$ is included in the rectangle of vehicle $i$ if $0 \leq x_j' \leq \sigma$ and $-w/2 \leq y_j' \leq w/2$; where $x_j'$ and $y_j'$ are calculated via:

$$\begin{aligned} x_j' &= (x_j - x_i)\cos\theta_i + (y_j - y_i)\sin\theta_i \\ y_j' &= (y_j - y_i)\cos\theta_i - (x_j - x_i)\sin\theta_i \end{aligned} \quad (41)$$

The procedure must then be repeated by transforming the variables into new coordinates whose origin and angle are $(x_j, y_j)$ and $\theta_j$, respectively.

## APPENDIX B: TRANSFORMED ISO-DISTANCE CURVES

For the iso-distance curve transformation to align with the desired orientation, new polar coordinates are defined whose origin is found by crossing two lines perpendicular to the vehicles' rear axles, while having the ego vehicle's desired deviation from the circular angle. This implies that the coordinates are different for each pair of vehicles. The lines' equations, according to Fig. 20, can be written as

$$\begin{aligned} y &= \tan(\varphi_i + s_{d,i})(x - r_i\cos\varphi_i) + r_i\sin\varphi_i \\ y &= \tan(\varphi_j + s_{d,i})(x - r_j\cos\varphi_j) + r_j\sin\varphi_j \end{aligned} \quad (42)$$

Note that each vehicle projects other vehicles on its own coordinates; therefore, in both equations of (42), the desired orientation of vehicle $i$ is taken into account. Intersecting above

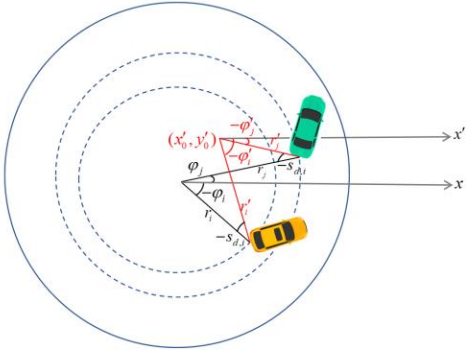

**Fig. 20.** Vehicles' variables in the original and new coordinates

equations gives the centre of new coordinates as

$$x_0' = \frac{r_i\left(\cos\varphi_i \tan(\varphi_i + s_{d,i}) - \sin\varphi_i\right)}{\tan(\varphi_i + s_{d,i}) - \tan(\varphi_j + s_{d,i})}$$
$$- \frac{r_j\left(\cos\varphi_j \tan(\varphi_j + s_{d,i}) - \sin\varphi_j\right)}{\tan(\varphi_i + s_{d,i}) - \tan(\varphi_j + s_{d,i})} \quad . \quad (43)$$
$$y_0' = \tan(\varphi_i + s_{d,i})(x_0' - r_i\cos\varphi_i) + r_i\sin\varphi_i$$

Then, the radius and angle of each vehicle in the new coordinates is calculated by

$$\begin{aligned} r' &= \sqrt{(x-x_0')^2 + (y-y_0')^2} \\ &= \sqrt{(r\cos\varphi - x_0')^2 + (r\sin\varphi - y_0')^2} \\ \varphi' &= \tan^{-1}\left((y-y_0')/(x-x_0')\right) \\ &= \tan^{-1}\left((r\sin\varphi - y_0')/(r\cos\varphi - x_0')\right) \end{aligned} \quad . \quad (44)$$

Furthermore, the distance in the new coordinates is

$$D_{i,j}' = \sqrt{p(r_i' - r_j')^2 + 2r_i'r_j'(1-\cos(\varphi_i' - \varphi_j'))} \quad (45)$$

According to (42), the centre of new coordinates is on the perpendicular line of the ego vehicle. If the centre is located between the ego and adjacent vehicle, which happens in a very limited area around the ego vehicle, the transformed distance $D_{i,j}'$ dramatically drops. To avoid this, the ego vehicle position is considered as the centre in this situation.

## APPENDIX C: LOCAL DENSITY

To determine the local density, a moving rectangle with length $L$ and width $W$ around the vehicle is considered which is aligned with the ego vehicle's orientation, see Fig. 21. The parameter $\eta \in [0,1]$ determines the ratio of the covered area in front of the vehicle to the whole rectangle area. Specifically, if $\eta = 1$, the rectangle only considers the vehicles in front. Then, the local density $\rho$ is defined as the portion of the rectangle area covered by vehicles, excluding the ego vehicle. When a vehicle is not completely in the rectangle, a fraction of its area which is within the rectangle is considered. Also, only the part of the rectangle which is within the road boundaries is considered for local density calculation.

## APPENDIX D: SAFETY CONTROLLER DETAILS

### A. Finding cross-point when EV and obstacle have circular and skewed motions, respectively

In this case, the cross-point can be found by crossing the following movement equations:

$$\begin{aligned} y &= \pm\sqrt{r_i^2 - x^2} \\ y &= \tan\theta_{d,i}(x - x_j) + y_j \end{aligned} \quad (46)$$

So, the cross-point's longitudinal position is one of the roots of the following second-order polynomial that results from (46) after simplification:

$$\begin{aligned} ax_c^2 + bx_c + c &= 0 \\ a &= 1 + \tan^2\theta_{d,j} \\ b &= 2(\tan\theta_{d,j} - x_j\tan^2\theta_{d,j}) \\ c &= x_j^2\tan^2\theta_{d,j} - 2x_j\tan\theta_{d,j} + y_j^2 - r_i^2 \end{aligned} \quad (47)$$

If $\Delta = b^2 - 4ac < 0$, the roots are imaginary, that is, the motions do not cross. If $\Delta = 0$, there is one single root, which will be considered as the cross-point. If $\Delta > 0$, there are two real roots for which the corresponding lateral position and angle in the polar coordinates must be calculated. The lateral position, then, can be determined by replacing $x_c$ in (46). A root can be considered as a candidate if its angle is downstream both EV and obstacle. If both roots have this characteristic, then the closer one is considered as the cross-point.

### B. Finding cross-point when EV and obstacle have skewed and circular motions, respectively

Following a similar procedure to the previous one, the cross-point can be found by calculating roots of the following equation:

$$\begin{aligned} ax_c^2 + bx_c + c &= 0 \\ a &= 1 + \tan^2\theta_{d,i} \\ b &= 2(\tan\theta_{d,i} - x_i\tan^2\theta_{d,i}) \\ c &= x_i^2\tan^2\theta_{d,i} - 2x_i\tan\theta_{d,i} + y_i^2 - r_j^2 \end{aligned} \quad (48)$$

The found cross-points should be transformed to the coordinates aligned with the EV's desired orientation. If they are behind the EV or obstacle or the roots are not real, they are ignored. Otherwise, the closest cross-point is considered.

### C. Finding the closest current obstacle

In the case of circular EV motion, being near the path of EV

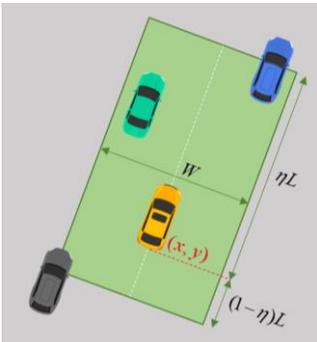

**Fig. 21.** Local density rectangle

means having a larger angle in polar coordinates ($0 \leq \varphi_j - \varphi_i < \pi/2$) and a radius close to the ego vehicle radius ($|r_j - r_i| < w + w_{th}$) where $w_{th}$ is a safety margin (e.g. equal to 2 m). Similarly, for skewed motions, it is defined as having a bigger longitudinal position in the transformed coordinates aligned with the vehicle orientation ($x'_j \geq x'_i$), while having a lateral position sufficiently close ($|y'_j - y'_i| < w + w_{th}$). The transformed variables can be calculated by (6), while the EV's current direction is considered as $\theta'$. Among the obstacles located in the mentioned area, the closest one to the EV is taken as the closest obstacle.

### APPENDIX E: CONTROLLER PARAMETERS

The local density and adaptive desired speed parameters are $L = 80$ m, $W = 10$ m, $\eta = 1$, $\sigma_{safety} = 3.5$ m, $w_{safety} = 2$ m, $\lambda_s = 1.3$, $\lambda_r = 0.02$; in the safety controller, $w_{th} = 2$ m, $D_s = 7$ m; also, $D_0 = 5$ m and $D_1 = 6$ s in circular part ($D_0 = 10$ m and $D_1 = 8$ s in straight roads). The feedback gains used in the boundary and safety controllers are: $\mathbf{K} = [1.5 \quad 1.9]$, $\mathbf{K}_c = [-52 \quad 46]$, and $\mathbf{K}_s = [20 \quad 9]$. Also, $\varphi_b$, used in the formation of corridors for invisible ODs, is $30°$.

Although several parameters in both straight and circular NLFCs are noted with the same symbol, they may have different values, as their roles differ. In addition, some parameters have different values in different phases, as mentioned in Section III. In straight NLFC (9), $A = 0.5$, $\varepsilon = 0.1$, $p = 1.5$, $\gamma_1 = 0.02 + 1.1v$, $\gamma_2 = 4$, and $\gamma_3 = 9$. For entering vehicles $\mu_1 = 0.3$, $\mu_2 = 0.1$, and $\Theta = 10°$ while for exiting vehicles $\mu_1 = 3$, $\mu_2 = 7$, and $\Theta$ is $80°$; in circular NLFC (13), $A = 0.005$, $b = 1.2$, $\lambda = 25$ m, $q = 0.02$, $\varepsilon = 0.1$, $p = 3$, $\gamma_1 = 0.0004 + 0.03v$, $\mu_1 = 10$, in the entering and exiting phases $\gamma_2 = 3.5$, $\mu_2 = 80$ and $\Theta = 80°$ while in the rotation phase, $\gamma_2 = 6$, $\mu_2 = 40$, and $\Theta = 50°$. In both controllers $v_{max}$ is $25$ m/s. Finally, to distribute vehicles properly on the roundabout surface, the averaging weight $\alpha$ in (27) is randomly chosen in the range $[0.2, 0.55]$ for each vehicle.

The choice of most of these parameters was straightforward, based on their functionality, as explained in previous sections. For a few of them, some fine-tuning was carried out using preliminary simulations. The sensitivity of the reported results with respect to all design parameters is low.

15
> REPLACE THIS LINE WITH YOUR MANUSCRIPT ID NUMBER (DOUBLE-CLICK HERE TO EDIT) <[27] A. Koumpis, D. Zorzenon, and F. Molinari, "Automation of roundabouts via consensus-based distributed auctions and stochastic model predictive control", *IEEE conference on European Control Conference (ECC)*, London, United Kingdom, 2022, pp. 14-20.
[28] J.P. Rastelli, and M.S. Penas, "Fuzzy logic steering control of autonomous vehicles inside roundabouts", *Applied Soft Computing*, 35, 2015, pp. 662-669.
[29] N. Ding, X. Meng, W. Xia, D. Wu, L. Xu, and B. Chen, "Multi-vehicle coordinated lane change strategy in the roundabout under internet of vehicles based on game theory and cognitive computing", *IEEE Transactions on Industrial Informatics*, 14(8), 2015.
[30] P. Hang, C. Huang, Z. Hu, Y. Xing, and C. Lv, "Decision making of connected automated vehicles at an unsignalized roundabout considering personalized driving behaviours", *IEEE Transactions on Vehicular Technology*, vol. 70, no. 5, 2021.
[31] R. Tian, S. Li, N. Li, I. Kolmanovsky, A. Girard, and Y. Yildiz, "Adaptive game-theoretic decision making for autonomous vehicle control at roundabouts", *IEEE Conference on Decision and Control (CDC)*, 2018, pp. 321-326.
[32] Y. Zhang, B. Gao, L. Guo, H. Guo, and H. Chen, "Adaptive decision-making for automated vehicles under roundabout scenarios using optimization embedded reinforcement learning", *IEEE Transactions on Neural Networks and Learning Systems*, 32(12), 2020, pp.5526-5538.
[33] B. Németh, Z. Farkas, Z. Antal and P. Gáspár, "Hierarchical control design of automated vehicles for multi-vehicle scenarios in roundabouts," *IEEE European Control Conference (ECC)*, London, United Kingdom, 2022, pp. 1964-1969
[34] M. Naderi, M. Papageorgiou, I. Karafyllis, I. Papamichail, "Automated vehicle driving on large lane-free roundabouts", *25th IEEE International Conference on Intelligent Transportation Systems (ITSC)*, Macao, China, 2022, pp. 1528-1535.
[35] M. Naderi, M. Mavroeidi, I. Papamichail, M. Papageorgiou, "Optimal orientation for automated vehicles on large lane-free roundabouts", *IEEE Conference on Decision and Control (CDC),* 2023, accepted.
[36] P. Polack, F. Altche, B. d'Andrea-Novel, and A. de La Fortelle, "The kinematic bicycle model: A consistent model for planning feasible trajectories for autonomous vehicles?", *IEEE Intelligent Vehicles Symposium (IV)*, 2017, pp.812-818.
[37] D. Theodosis, F.N. Tzortzoglou, I. Karafyllis, I. Papamichail, and M. Papageorgiou, "Sampled-data controllers for autonomous vehicles on Lane-Free Roads", *30th Mediterranean Conference on Control and Automation (MED),* Athens, Greece, 2022, pp. 103-108.
[38] I. Karafyllis, D. Theodosis, M. Papageorgiou, "Analysis and control of a non-local PDE traffic flow model", *International Journal of Control,* 95, 2022, pp. 660-678.
[39] I. Karafyllis, D. Theodosis, M. Papageorgiou, "Constructing artificial traffic fluids by designing cruise controllers", *Systems & Control Letters,* 167, 2022, 105317.
[40] D. Troullinos, G. Chalkiadakis, D. Manolis, I. Papamichail, M. Papageorgiou, "Extending SUMO for lane-free microscopic simulation of connected and automated vehicles", *SUMO Conference Proceedings*, 2022, pp. 95:103.
[41] P. A. Lopez *et al*., "Microscopic traffic simulation using SUMO", 2018 *21st IEEE International Conference on Intelligent Transportation Systems (ITSC)*, Maui, HI, USA, 2018, pp. 2575-2582.

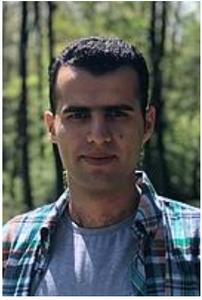
**Mehdi Naderi** received the B.S. degree in Electrical Engineering from the Hamedan University of Technology, Hamedan, Iran, in 2010, the M.Sc. degree in Electrical Engineering (Control) from the University of Tehran, Tehran, Iran, in 2013, and the Ph.D. degree in Electrical Engineering (Control) from the K.N. Toosi University of Technology (KNTU), Tehran, Iran, in 2019. From 2020 to 2021, he was a lecturer at KNTU. His research interests include autonomous vehicles, control allocation, and fault tolerant control systems. Since May 2021, he has been a post-doc associate with the Dynamic Systems & Simulation Laboratory, Technical University of Crete.

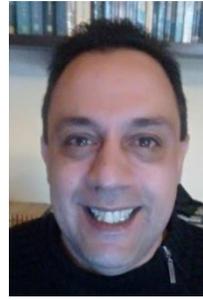
**Iasson Karafyllis** is a Professor in the Department of Mathematics, NTUA, Greece. He is a coauthor (with Z.-P. Jiang) of the book Stability and Stabilization of Nonlinear Systems, Springer-Verlag London, 2011 and a coauthor (with M. Krstic) of the books Predictor Feedback for Delay Systems: Implementations and Approximations, Birkhäuser, Boston 2017 and Input-to-State Stability for PDEs, SpringerVerlag London, 2019. Since 2013 he is an Associate Editor for the International Journal of Control and for the IMA Journal of Mathematical Control and Information. Since 2019 he is an Associate Editor for Systems and Control Letters and Mathematics of Control, Signals and Systems. His research interests include mathematical control theory and nonlinear systems theory.

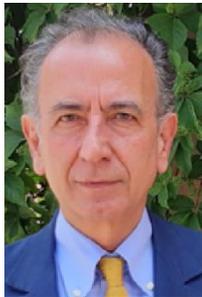
**Markos Papageourgiou** (Life Fellow, IEEE) was a Professor of Automation with the Technical University of Munich, Germany, from 1988 to 1994. Since 1994 he has been a professor (since 2021 Professor Emeritus) at the Technical University of Crete, Chania, Greece. Since 2021 he has been a Professor at Ningbo University, China. He was a Visiting Professor with the Politecnico di Milano, the Ecole´ Nationale des Ponts et Chaussées, MIT, the Sapienza University of Rome, and Tsinghua University, and a Visiting Scholar with UC Berkeley. His research interests include automatic control and optimisation theory and applications to traffic and transportation systems, water systems, and further areas. He is a fellow of IFAC. He received several distinctions and awards, including the 2020 IEEE Transportation Technologies Award and two ERC Advanced Investigator Grants.

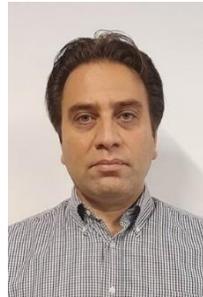
**Prof. Ioannis Papamichail** is the Director of the Dynamic Systems and Simulation Laboratory, at the Technical University of Crete, Chania, Greece. He received the Dipl. Eng. degree in chemical engineering from the National Technical University of Athens, in 1998 and the M.Sc. degree in process systems engineering and the Ph.D. degree in chemical engineering from Imperial College London, in 1999 and 2002, respectively.
From 1999 to 2002, he was a Research Assistant with the Center for Process Systems Engineering, Imperial College London. He joined the Technical University of Crete in 2004 and has served, since then, at all academic ranks. In 2010, he was a Visiting Scholar with the University of California, Berkeley, CA, USA. He is the author of several technical papers in scientific journals and conference proceedings. His main research interests include automatic control and optimization theory and applications to traffic and transportation systems.
Dr. Papamichail is an Associate Editor for Transportation Research Part C: Emerging Technologies and for IEEE Transactions on Intelligent Transportation Systems. He received the 1998 Eugenidi Foundation Scholarship for Postgraduate Studies and the 2010 Transition to Practice Award from the IEEE Control Systems Society for the development and implementation of ramp metering algorithms, particularly at the Monash Freeway, Melbourne, Australia. He also received the TRA 2012 Best Paper Award for Pillar II (Transport, Mobility and Infrastructure), the Best Freeway Operations Paper in 2014 Award by the Transportation Research Board (TRB) Freeway Operations Committee, and the IEEE-ITS 2020 Best Conference Paper Award.

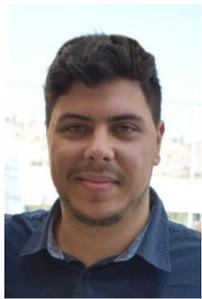
**Dimitrios Troullinos** received the Diploma degree in Electrical and Computer Engineering from the Technical University of Crete, Chania, Greece, in October 2019. His main research interests lie in the area of Multiagent Systems. From November 2019, he is a research associate and since January 2020 he has been a PhD student with the Dynamic Systems & Simulation Laboratory, Technical University of Crete.